\documentclass[reqno]{amsart}
\usepackage{graphicx,amsmath}
\usepackage{float}
\usepackage[utf8]{inputenc}
\usepackage{color}
\usepackage{cite}
\usepackage{fancyhdr}

\begin{document}
\pagestyle{plain}
\title{When theories and experiments meet: rarefied gases as a benchmark of non-equilibrium thermodynamic models}

\author{R. Kovács$^{123}$, P. Rogolino$^4$, D. Jou$^5$}

\address{
$^1$Department of Energy Engineering, Faculty of Mechanical Engineering, BME, Budapest, Hungary 
$^2$Department of Theoretical Physics, Wigner Research Centre for Physics,
Institute for Particle and Nuclear Physics, Budapest, Hungary
$^3$Montavid Thermodynamic Research Group
$^4$Department of Mathematics and Computer Sciences, Physical Sciences and Earth Sciences, University of Messina, Messina, Italy
$^5$Universitat Aut\'noma de Barcelona, Departament de Física, Bellaterra, Catalonia, Spain}

\date{\today}

\begin{abstract}
The role of thermodynamics in deriving constitutive equations is unique, and various approaches have been developed in the last decades. In the present paper, the frameworks of Extended Irreversible Thermodynamics (EIT) and Non-Equilibrium Thermodynamics with Internal Variables (NET-IV) are discussed and compared to each other on the basis of a particular problem of rarefied gases. In this comparison, both theoretical and experimental aspects are taken into account. Eventually, an experiment by Meyer and Sessler covering a wide range of pressures and frequencies is investigated. Here, concentrating on the scaling properties and the density dependence of parameters, the change of speed of sound in terms of frequency and pressure is recovered using NET-IV, and this fitting is compared to the results of Lebon and Cloot using EIT. 

\end{abstract}
\maketitle

\section{Introduction}

The generalization of the classical constitutive equations for heat, mass, charge and momentum transport is a longlasting task in non-equilibrium thermodynamics. Many approaches evolved in this direction, which can be compatible with each other under various conditions. In this paper, particularly the frameworks of Extended Irreversible Thermodynamics (EIT) and Non-Equilibrium Thermodynamics with Internal Variables (NET-IV) are compared in regard to modeling ultrasound propagation in rarefied gases. The earlier results of Lebon, Cloot, Jou et al. \cite{LebCloo89, JouEtal85, JouEtal10b, Lebon89, Lebon14, JouEtal99, JouVasLeb88ext, JouRes11, JouRes16, Restuccia16, RogCimm19} using EIT show not only possible extensions of the classical Navier-Stokes-Fourier (NSF) system but certain evaluation of experimental results as well. These experiments were performed by Meyer and Sessler \cite{MeySess57} and are in agreement with other works of Rhodes, Greenspan, Sluijter et al. \cite{Rhod46, SetEtal55, Gre56, SluiEtal64, SluiEtal65}, namely, a significant change in the speed of sound is observed for different frequencies and pressures, in particular, for different frequency-pressure ratios. 

The approach of EIT takes the heat flux and the viscous pressure as additional independent variables in a generalized entropy density and entropy flux, and it complements the microscopic results with some aspects from higher-order kinetic theory, especially the expressions of transport coefficients, i.e., the viscosities, and the thermal conductivity. However, the derivation of constitutive equations is, in several aspects, similar to NET-IV: calculating the entropy production and solving the resulting inequality leads to the constitutive laws. As such, EIT is somewhere between the Rational Extended Thermodynamics (RET) where the mathematical structure of the macroscopic evolution equations for higher-order moments is taken analogous to that arising from the microscopic kinetic theory, \cite{MulRug98, RugSug15}, and NET-IV which insists on purely phenomenological grounds. 

The rarefied gas analysis using NET-IV has been developed only recently \cite{KovEtal18rg, Kov18rg}. In this framework, EIT occurs as a particular case of NET-IV when the internal variables are identified to be the dissipative currents. However, in NET-IV, there is the possibility to be phenomenological, and there are no constraints either for the coupling among different variables or for the transport coefficients \cite{KovVan15}. As a result, it offers more degrees of freedom in the modeling task, and examples can be found in \cite{Kov18rg, KovVan16, KovVan18} where both heat conduction and rarefied gas experiments are discussed. 

In the following, the paper presents the fundamental differences between the approaches of EIT and NET-IV (Section 2), and the respective resulting systems of generalized NSF equations. After concluding the theoretical results, a particular experiment on ultrasound propagation in rarefied gases is evaluated (Section 3), and the related modeling aspects will be summarized and discussed (Section 4).

\section{Generalization of constitutive equations}

In order to derive the classical NSF equations, the internal energy $e$ and the mass density $\rho$ are enough as variables of entropy density: $s=s(e,\rho)$ together with the classical entropy flux $\mathbf J_s=\mathbf q/T$ with $T$ being the absolute temperature and $\mathbf q$ standing for the heat flux. The required balance equations are the mass, momentum and energy, namely,
\begin{align}
\dot \rho + \rho \nabla \cdot \mathbf v &=0, \label{massbal} \\
\rho \dot {\mathbf v} + \nabla \cdot \mathbf P &=0, \label{mombal} \\
\rho \dot e + \nabla \cdot \mathbf q &= - \mathbf  P : \nabla \mathbf v, \label{enerbal}
\end{align}
respectively. Here, the upper dot represents the material time derivative, $\mathbf P$ is the pressure tensor and $\mathbf v$ is the local barycentric velocity, and the double dot stands for double contraction of the corresponding tensors. One can formulate the second law of thermodynamics as an inequality,
\begin{align}
\rho \dot s + \nabla \cdot \mathbf J_s = \sigma_s \geq 0, \label{entin}
\end{align}
in which the entropy production $\sigma_s$ consists of a sum of products of thermodynamic forces and fluxes according to Onsager \cite{Onsager31I, Onsager31II}. As a solution of eq.~(\ref{entin}), the Newton's law for fluids is obtained that connects the viscous pressure $\mathbf \Pi=\mathbf P-p \mathbf I$ ($\mathbf I$ is the identity tensor and $p$ the static pressure) to the velocity gradient $\nabla \mathbf v$, i.e.,
\begin{align}
\mathbf \Pi_d=-\nu (\nabla \mathbf v)_d, \quad \Pi_s=-\eta \nabla \cdot \mathbf v,
\end{align}
where $\nu$ and $\eta$ are the shear and volumetric viscosities, respectively, and the indices $d$ and $s$ are for distinguishing the deviatoric and spherical parts. The Curie principle ensures that the Fourier's law is not coupled to the pressure at the linear level of constitutive equations, that is,
\begin{align}
\mathbf q= - \lambda \nabla T,
\end{align}
with $\lambda$, the thermal conductivity forms separately from scalar and tensorial fluxes and forces. 

EIT proposes to extend the space of independent variables with dissipative fluxes such as the heat flux and the viscous pressure $\mathbf \Pi$. Applying the classical entropy flux, it results in hyperbolic evolution equations, see Gyarmati \cite{Gyar77a, Gyarmati70b} for details. This is one way to derive the famous Maxwell-Cattaneo-Vernotte equation \cite{Max1867, Cattaneo58, Vernotte58} which is applicable for second sound and introduces a sort of inertia of heat conduction. Introducing the viscous pressure instead of the heat flux as an independent variable, gives Meixner's theory \cite{Meix43a, GrooMaz63non} of viscous relaxation. Working with NET-IV, all these results can be recovered as a special case when the internal variables \cite{Verhas97, BerVan17p, BerVan17b, SzucsFul19, SzucsFul18b, Fama2019, CiaRes16} are identified as being these fluxes. Otherwise, it keeps the modeling on a more general level, and offers a system of coupled evolution equations but with more degrees of freedom. 

The other difference between EIT and NET-IV emerges in the formulation of entropy flux. In EIT, $J_s$ is generalized using products of fluxes and gradients of fluxes that creates coupling between vectors and tensors in a natural way. Applying NET-IV, it can be achieved using the Nyíri (or current) multipliers \cite{Nyiri89, Nyiri91} which are known to be constitutive functions; however, with unspecified form. The entropy inequality yields it as a solution, too. 
In the present paper, only the relevant equations are presented which are used for rarefied gas modeling. 

\subsection{EIT model for rarefied gases}
Summarizing the work of Lebon and Cloot \cite{LebCloo89} on sound propagation at high frequencies in rarefied gases, that was preceded by the analysis of Carrasi and Morro \cite{CarMorr72a, CarMorr72b}, the heat flux $\mathbf q$ and the traceless part of the viscous pressure $\mathbf \Pi_d$ are introduced besides the internal energy and the mass density as state variables of the extended entropy density s, that is, $s=s(e,\rho,\mathbf q, \mathbf\Pi_d)$. It is worth emphasizing that only the deviatoric part of the pressure is used and the spherical part is omitted in their analysis, because they assume dilute monatomic gases without internal degrees of freedom.
The extended entropy current reads
\begin{align}
\mathbf J_s=\mathbf q/T+\alpha_1 \mathbf \Pi_d \cdot \mathbf q+\alpha_2 \mathbf q \cdot \nabla \mathbf q + \alpha_3 \mathbf \Pi_d : \nabla \mathbf \Pi_d, \label{eqjseit}
\end{align}
where the first term is the classical one and the coefficients $\alpha_n$ (n=1,2,3) are phenomenological ones. In the simplest version of EIT, one takes $\alpha_2=\alpha_3=0$. 
Then calculating the entropy production $\sigma_s$ using eqs.~(\ref{entin}) and (\ref{eqjseit}), and solving the resulting inequality ($\sigma_s \geq 0$), the constitutive equations are
\begin{align}
\tau_q \dot{\mathbf q} &= - \mathbf q - \lambda \nabla T  - \lambda_2 \nabla \cdot \mathbf \Pi_d - \lambda_3 \Delta \mathbf q, \nonumber \\
\tau_d \dot{\mathbf \Pi}_d &=-\mathbf\Pi_d - \nu (\nabla \mathbf v)_d - \nu_2 (\nabla \mathbf q)_d - \nu_3 \Delta \mathbf\Pi_d \label{EITeq}
\end{align}
with $\Delta$ being the Laplacian, and $\tau_q$ and $\tau_d$ are the corresponding relaxation times of $\mathbf q$ and $\mathbf \Pi_d$. Together with the corresponding balances (\ref{massbal})-(\ref{enerbal}), and the equations of state
\begin{align}
p=\rho R T, \quad e=cT,
\end{align}
\eqref{EITeq} forms a closed system for $T$, $\rho$, $\mathbf v$, $\mathbf q$ and $\mathbf \Pi_d$. It is important to note that using such an extended entropy current density, the resulting equations are parabolic due to the Laplacian terms on the right hand side in (\ref{EITeq}). The Laplacian terms in \eqref{EITeq} arise from the terms in $\alpha_2$ and $\alpha_3$ in \eqref{eqjseit}, and they vanish for $\alpha_2=\alpha_3=0$. Indeed, it may be shown that $\lambda_3$ and $\nu_3$ are proportional to $\alpha_2$ and $\alpha_3$, respectively.

In order to preserve some compatibility with the kinetic theory, and to reduce the number of parameters appearing in equations \eqref{EITeq}, the following relations among the unknown 8 coefficients are assumed to be valid \cite{JouEtal10b}:
\begin{align}
\tau_q= \frac{3}{2} \tau_d, \quad \lambda=\frac{15}{4}R p \tau_d, \quad \lambda_2=\frac{3}{2}R T \tau_d, \nonumber \\
\nu=2 p \tau_d, \quad \nu_2=\frac{4}{5} \tau_d, \quad \nu_3=-RT\tau_d^2. \label{parrel}
\end{align}
Note that these relations, arising from second-order approximation in the Grad approach to kinetic theory \cite{JouEtal10b}, reduce the $8$ independent coefficients in \eqref{EITeq} to $2$ independent coefficients, $\tau_d$ and $\lambda_3$. In the final discussion, we will further comment on this point.
In the analysis of Lebon and Cloot, the two remaining parameters ($\lambda_3$ and $\tau_d$) are to be fitted with experimental results. Interestingly, the ratio of relaxation times $\tau_q$ and $\tau_d$ is fixed in \eqref{parrel}, and all coefficients are represented using $\tau_d$ except for $\lambda_3$. This is not usual in the framework of RET. 

\subsection{NET-IV model for rarefied gases}
In contrast to the previous EIT model, the spherical part of the viscous pressure is also considered as a state variable, i.e., $s=s(e,\rho,\mathbf q,\mathbf \Pi_d,\Pi_s)$. This allows to deal with the influence of internal degrees of freedom, as rotational and vibrational degrees of freedom in diatomic molecules.
 Moreover, in order to achieve the coupling between the thermal and fluid equations, the entropy current density $\mathbf J_s$ is generalized accordingly:
\begin{align}
\mathbf J_s=(\mathbf b_d + b_s \mathbf I) \mathbf q,
\end{align}
where $\mathbf b_d$ and $b_s$ are the deviatoric and the spherical parts of the Nyíri-multiplier. This is called `nonlocal generalization' of entropy density and its current density \cite{Kov18rg}. The resulting constitutive relations in one spatial dimension are
\begin{align}
\tau_q\partial_t q +q +\lambda \partial_x T - \alpha_{21} \partial_x \Pi_s - \beta_{21} \partial_x \Pi_d =& 0, \nonumber \\
\tau_d \partial_t \Pi_d +\Pi_d + \nu \partial_x v + \beta_{12} \partial_x q =& 0, \nonumber \\
\tau_s \partial_t \Pi_s +\Pi_s + \eta \partial_x v + \alpha_{12} \partial_x q =& 0,
\label{NETSYSLINFINAL}
\end{align}
where $\alpha_{ab}$, $\beta_{ab}$ ($a,b=1,2$) are the coupling parameters and $\tau_m$ ($m=q,d,s$) are the corresponding relaxation times. Its three dimensional derivation and for detailed calculation, see \cite{KovEtal18rg}. In summary, there are 7 parameters to be fit in eq.~\eqref{NETSYSLINFINAL}, the classical transport coefficients are considered to be known. Note that in \eqref{NETSYSLINFINAL}, the Laplacian terms $\Delta \mathbf q$ and $\Delta \mathbf \Pi_d$ from \eqref{EITeq} do not appear.
This model is compatible with the results of Arima et al.~\cite{Arietal12, AriEtal12c, Arietal13, Arietal15}.

\subsection{Remarks and comparison}
It is well-known that the derivation of constitutive equations is one common point between EIT and NET-IV. However, the differences originate in the form of entropy flux and the interpretation of coefficients. The framework of NET-IV is more phenomenological, proposes no direct connection among the relaxation times and the other transport coefficients. Nevertheless, EIT is compatible with NET-IV in the derivation method, and also compatible with kinetic theory concerning the formulation of coefficients through the relations \eqref{parrel}.  It decreases the number of free parameters significantly, from 8 to 2.

The first, successful evaluation of rarefied gas experiments is related to Arima et al. \cite{Arietal12, Arietal13}. They used the framework of RET which could be equivalent with NET-IV and EIT under certain conditions \cite{KovEtal18rg}. The corresponding linearized, one-dimensional equations are
\begin{align}
\tau_q \partial_t q + q + \lambda \partial_x T - R T_0 \tau_q \partial_x \Pi_d + R T_0 \tau_q \partial_x \Pi_s= &0, \nonumber \\
\tau_d \partial_t \Pi_d + \Pi_d + 2 \nu \partial_x v - \frac{2 \tau_d}{1+c_v^*}\partial_x q=&0, \nonumber \\
\tau_{s} \partial_t \Pi_s + \Pi_s + \eta \partial_x v + \frac{\tau_s (2c_v^*-3)}{3c_v^*(1 + c_v^*)} \partial_x q =&0, \label{ET_LINSYS2}
\end{align}
with $R$ being the gas constant and $c_v^*$ is the dimensionless specific heat: $c_v^*=c_v / R$. In this model, the transport coefficients are formulated in a different way than in the work of Lebon and Cloot, that is,
\begin{align}
\lambda=(1+c_v^*)R^2 \rho_0 T_0 \tau_q, \quad \mu=R\rho_0T_0\tau_p, \quad \eta=\left( \frac{2}{3} -\frac{1}{c_v^*}\right )R \rho_0 T_0 \tau_s, \label{param2}
\end{align}
where all three relaxation times are included into different transport coefficients. 
Note that for monatomic ideal gases without internal degrees of freedom, $c_v^*=3/2$, and as a consequence the coefficient $\eta$ in \eqref{param2}, the bulk viscosity, is zero, also appearing in the work of Lebon and Cloot.
Moreover, the derivation of these equations in RET is more complicated, requires a bi-hierarchy of balances but ensures Galilean invariance and thermodynamically compatible closure. For detailed discussion, the reader is invited to read \cite{KovEtal18rg}.

As previously shown, various theoretical approaches exist for the same problem. Actually, the structure of the constitutive equations is practically the same but they differ in their coefficients. On one hand, it restricts the validty (or the generality) of a model. On the other hand, it could ease or make more difficult the fitting procedure. In this sense, the most restricted model is EIT since this approach fixes almost all the coefficients by means of relations in \eqref{parrel}. The identification \eqref{parrel} from kinetic theory are incorporated. In fact, adopting of relations in \eqref{parrel} is not a thermodynamic result, but a complementary ansatz. In purely thermodynamic terms, all the coefficients in \eqref{parrel} should be fitted to experimental results. Since this would be too complicated, relations \eqref{parrel} are adopted.
The RET model leaves one more parameter to be fit, and interprets the meaning of coefficients in a different way. The most general and hence the most difficult model to fit is NET-IV since there is no any direct connection to the kinetic theory and as such, all the parameters (\# 7) are to be fit. 

One last remark must be made about the scaling properties of the models. Using EIT and RET, the frequency - pressure ratio appears naturally by a priori assuming an interaction model, nevertheless, such scaling requires constant viscosities and thermal conductivity (namely, independent of $\rho$, but not on $T$), and $1/\rho$ dependence in the relaxation times and also in the coupling parameters. In NET-IV, that question can be answered from a different point of view. If such a scaling is required then it restricts the density dependence of coupling coefficients and of the relaxation times in the same way ($\sim 1/\rho$). Naturally, constant transport coefficients are also required. However, as NET-IV keeps the modeling on a more general level, it allows to implement the density dependence of any parameter, even for the viscosities \cite{Kov18rg}. 

\section{Rarefied gas experiments}
In this section, the experiments performed by Meyer and Sessler \cite{MeySess57} are investigated. These are similar to the earlier measurements, conducted by Rhodes \cite{Rhod46} and Greenspan \cite{Gre56}. In these cases, the change in speed of sound has been measured in Argon gas with respect to the frequency-pressure ratio ($f/p$), at 20 $^\circ$C. There are two reasons why this particular experiment is chosen:
\begin{itemize}
\item there is an available evaluation using EIT \cite{LebCloo89},
\item the ratio of $f/p$ is especially high.
\end{itemize}
The latter one means that a wide interval is covered, from $10^7$ to $10^{11}$ Hz/atm. It points completely beyond the measurement of Rhodes where the maximum was around $10^7$ Hz/atm.

One recent evaluation for Rhodes' measurements \cite{Kov18rg} shows that in some cases, the frequency is given together with the temperature, hence the pressure (and the mass density, accordingly) can be calculated. In Rhodes' experiment, the pressure covers the range of $2-100$ kPa. In the experiment of Meyer and Sessler, it is $0.2-10000$ Pa ($f=100$ kHz). Since the pressure is varied within several orders of magnitude, the change in the transport coefficients could be important. It is experimentally shown that the viscosity tends to zero when the pressure is decreased and tends to zero \cite{IttPae40, ItterKee38, ItterCla38} for normal Hydrogen and it is expected to be similarly valid for any other gas. It is contradictory with the prediction from kinetic theory which tells non-zero viscosity at zero density \cite{TrueMunc80b}. Predictions of usual kinetic theory are valid when particle-particle collisions dominate, but not in the low-density domain such that particle-particle collision becomes irrelevant as compared to particle-wall collisions.
Moreover, the published zero-density limit viscosities are not measured\footnote{The papers of Gracki et al.~\cite{GrackiEtal69a, GrackiEtal69b} about the experimental data present exptrapolation from around $3$ kg/m$^3$ to zero.} but calculated \cite{AzizSlam86, MoldTrus88, MoldEtal99, MayEtal06}. Recalling the scaling properties of the theories, constant transport coefficients can be problematic in some region of interest; but serves as a requirement to obtain the $f/p$ scaling. Despite, the present analysis is restricted to constant viscosities (independent of $\rho$) in order to keep the compatibility with the other approaches. 

\begin{figure}[H]
\includegraphics[width=10cm,height=6cm]{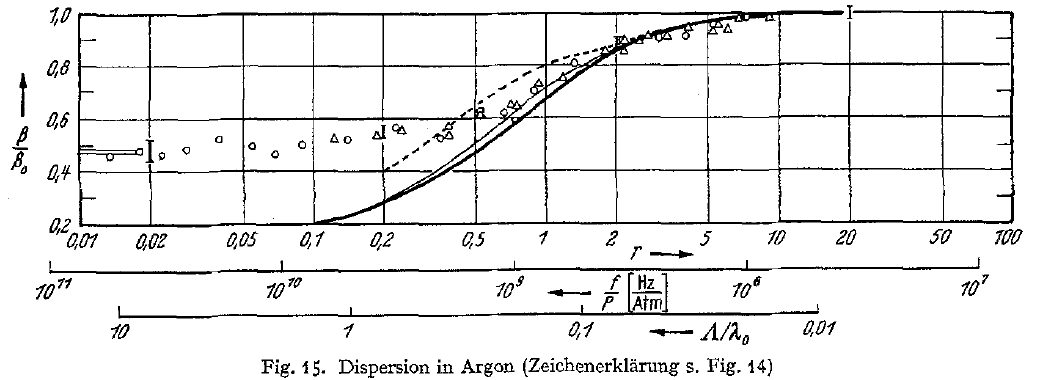}
\caption{Change in speed of sound in terms of the frequency - pressure ratio \cite{MeySess57}. Circles: measured data at $100$ kHz. Triangles: measured data at $200$ kHz. Thick line: classical theory. Thin line: Burnett theory. Dashed line: super-Burnett theory.}
\label{meysesmeas}
\end{figure}

In Fig.~\ref{meysesmeas}, the recorded experimental data are depicted \cite{MeySess57}showing a significant change in speed of sound. Furthermore, it also shows the evaluations using the classical, the Burnett and super-Burnett equations and all of them fail to model the behavior on high $f/p$ ratios. 

\begin{figure}[H]
\includegraphics[width=10cm,height=6cm]{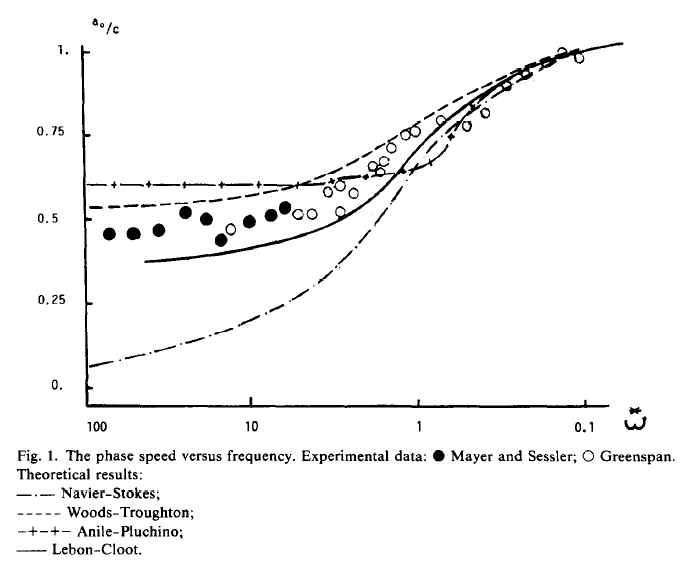}
\caption{Evaluation of the Meyer-Sessler experiment from EIT equations \eqref{EITeq} by Lebon and Cloot \cite{LebCloo89}.}
\label{meyseseval1}
\end{figure}

Fig.~\ref{meyseseval1} shows various evaluations, the most interesting one is drawn by the thick continous line, using EIT equations (\ref{EITeq}), by Lebon and Cloot \cite{LebCloo89}. Here the $f/p$ ratio is transformed to dimensionless frequency. Again, at high values, the theories under- or overpredict the phenomenon. 

\begin{figure}[H]
\includegraphics[width=10cm,height=7cm]{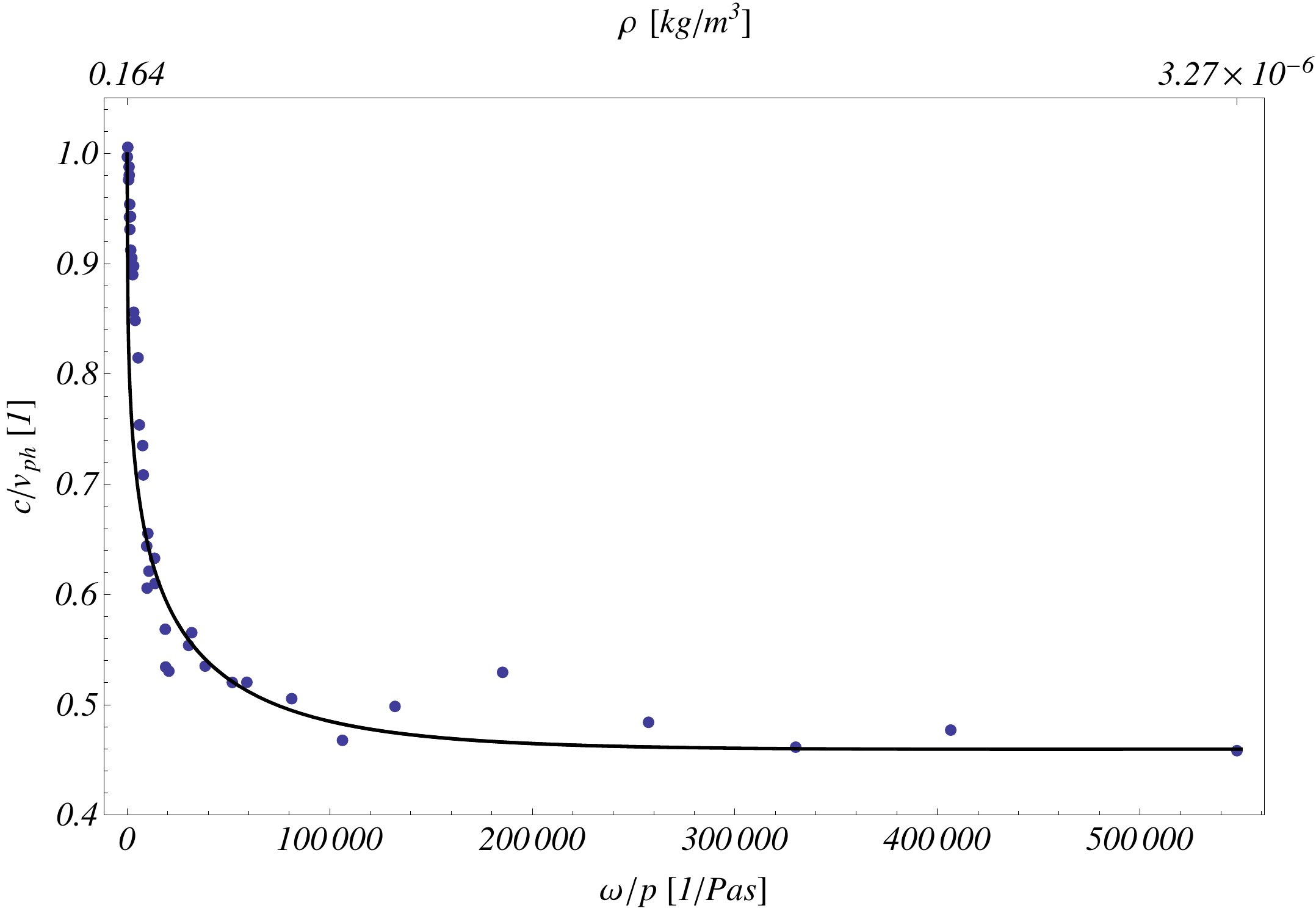}
\caption{Evaluation of the Meyer-Sessler experiment using NET-IV.}
\label{meyseseval2}
\end{figure}

In contrast to the previous results, Fig.~\ref{meyseseval2} shows a better fitting using the model of NET-IV (eq.~(\ref{NETSYSLINFINAL})). Thanks to the more degrees of freedom, it is possible to model the high $f/p$ behavior without any difficulty. However, it must be noted here that the density dependence of the classical transport coefficients is neglected which could be a significant source of errors in any models but here the higher number of free parameters can overcome this shortcoming. 

Another aspect to be taken into consideration is that the relaxation times of rotational and vibrational degrees of freedom, related to $\Pi_s$ and $\tau_s$, are not expected to depend on density or pressure in the same form as the relaxation times related to the particle-particle collisions. This could make that in some range of pressure, in which the rotational and vibrational degrees of freedom are comparatively relevant, the density dependence of velocity in terms of $f/p$ could be different from that predicted on the basis of translational degrees of freedom where $\tau_d$ and $\tau_q$ depend on $\rho$.

\begin{table}[H]
\caption{Fitted relaxation time coefficients for continuum model based on NET-IV.} \label{expcoeff1}
\centering
\begin{tabular}{ccc} $\tau_q = \frac{t_1}{\rho}$, $t_1 = \left [\frac{kg \cdot s}{m^3} \right ]$ & $\tau_d = \frac{t_2}{\rho}$,  $t_2 = \left[\frac{kg \cdot s}{m^3}\right]$ & $\tau_s = \frac{t_3}{\rho}$ $t_3 =\left [\frac{kg \cdot s}{m^3}\right]$\\ \hline
$3.05 \cdot 10^{-7}$ &	$1.55 \cdot 10^{-6}$	& $0.141 \cdot 10^{-9}$\\
\end{tabular}
\end{table}

\begin{table}[H]
\caption{Fitted coupling coefficients for continuum model based on NET-IV.} \label{expcoeff2}
\centering
\begin{tabular}{cc} $\alpha_{12}= \frac{a_{12}}{\rho}$, $a_{12} =\left [\frac{kg \cdot s}{m^3}\right]$ &  $\beta_{12}= \frac{b_{12}}{\rho}$, $b_{12} =\left [\frac{kg \cdot s}{m^3}\right]$  \\  $3.89 \cdot 10^{-7}$ &	$1.02 \cdot 10^{-4}$	 \\ \hline
$\alpha_{21}=\frac{a_{21}}{ \rho}$, $a_{21} =\left [\frac{kg}{m \cdot s}\right]$ & $\beta_{21}=\frac{b_{21}}{ \rho}$, $b_{21} =\left [\frac{kg}{m \cdot s}\right]$ \\ $6.23 \cdot 10^{-7} $ & $6.4 \cdot 10^{-4}$\\
\end{tabular}
\end{table}


\section{Discussion}
The basic differences between the frameworks of Extended Irreversible Thermodynamics and Non-Equilibrium Thermodynamics with Internal Variables are presented and both are tested on the same experiment on ultrasound propagation in dilute gases, which serves as a benchmark to comparing and checking the validity of thermodynamic theories. In case of EIT, relations \eqref{parrel} for coefficients are taken from kinetic theory. This is a simplifying ansatz which does not strictly follows from thermodynamics but from compatibility of second-order kinetic theory. There are only two parameters to fit ($\tau_d$ and $\lambda_3$ in \eqref{EITeq}), which is much easier than using NET-IV; nevertheless the outcome is less precise and less general. For certain $f/p$ region EIT seems to be an easier way to apply while its precision is enough. When a specific density dependence should be implemented, then the generality of NET-IV comes handy. Tables \ref{expcoeff1} and \ref{expcoeff2} summarize the fitted parameters for NET-IV. 

Indeed, taking only a few moments of the distribution function as variables - or, alternatively, a small number of fluxes - and neglecting the higher-order moment or higher-order fluxes has some subtle physical consequences \cite{JouEtal10b, LuzziEtal98}. In particular, the relaxation times appearing in \eqref{parrel} must be replaced by the mentioned renormalized effective relaxation times, which may be smaller than the relaxation times of simple kinetic theory in a factor between $1/2$ and $1/4$, approximately. Thus, the comparison of kinetic theory and observations is not a simple task at high frequencies. Keeping a phenomenological freedom may be useful and reasonable in this context.


Table \ref{compEN} compares some elementary aspects of these approaches. One of the outcome properties is the hyperbolicity property of the models. Applying EIT, the outcome can be influenced by assuming a proper entropy flux. In this paper, a parabolic system is presented that can be recognized by the Laplacian terms in the constitutive equations \cite{RogEtal17}. Moreover, there are other room temperature experiments, especially for heat conduction, where parabolic models performs significantly better \cite{Botetal16, Vanetal17, FulEtal18e}. Applying a current multiplier in the framework of NET-IV, always leads to parabolic models with richer structure than the corresponding models from RET. These parabolic equations can be simplified to hyperbolic ones which are in agreement with RET and leave the coefficients unconstrained. 
For instance, the coefficients $\alpha_2$ and $\alpha_3$ in the entropy flux \eqref{eqjseit} are zero, the equations become hyperbolic. This illustrates the relevance of the expression for the entropy flux in the form of equations.

As a benchmark, a particular experiment conducted by Meyer and Sessler \cite{MeySess57} is investigated in detail. Lebon and Cloot \cite{LebCloo89} performed its evaluation and found a better description at the high $f/p$ region than several other contemporary models. It is shown that NET-IV can improve the fitting on this domain, demonstrating that it covers the whole range without a seeming restriction. Such improvement could be due to accounting the spherical part of the viscous pressure tensor. Moreover, EIT and NET-IV share another common attribute: both are using phenomenological coefficients which offer higher level of generality. If needed, one may still assume a correspondance with the kinetic theory, but it is not a necessary requirement and does not influence the validity region of the approach.

\begin{table}[H]
\caption{Comparison of some relevant aspects of EIT and NET-IV}  \label{compEN}
\begin{tabular}{lcc} 
       & EIT    & NET-IV                                                                           \\ \hline
Input  & \begin{tabular}[c]{@{}c@{}}fluxes as new variables\\ gradient extensions in the entropy current\\ kinetic theory for the coefficients\end{tabular} & \begin{tabular}[c]{@{}c@{}}internal variables\\ current multipliers\end{tabular} \\ \hline
Output & \multicolumn{2}{c}{\begin{tabular}[c]{@{}c@{}}hyperbolic or parabolic equations\\ coefficients are phenomenological: fitted or adopted\end{tabular}}                                                                                 
\end{tabular}
\end{table}

\section{Acknowledgement}
\noindent R.K.: The research reported in this paper has been supported by the National Research, Development and Innovation Fund (TUDFO/51757/2019-ITM), Thematic Excellence Program. The work was supported by the grants of National Research, Development and Innovation Office – NKFIH, NKFIH K123815 and NKFIH KH130378. \\
D.J. acknowledges the financial support of the Spanish Ministry of Economy and Competitiveness under grant RTI 2018-097876-B-C22 and of the University of Messina as a visiting researcher in Messina in April and June 2019 (resolution of the Academic Senate, 23 July 2018, protocols 56199 and 56210).\\
The authors acknowledge the financial support of the Italian Gruppo Nazionale per la Fisica Matematica (GNFM-INdAM).


\end{document}